\documentclass[conference]{IEEEtran}
\IEEEoverridecommandlockouts
\usepackage{cite}
\usepackage{amsmath,amssymb,amsfonts}
\usepackage{graphicx}
\usepackage{textcomp}
\usepackage{xcolor}
\usepackage{booktabs}
\usepackage{array}
\usepackage{comment}
\usepackage[hyphens]{url}

\IEEEaftertitletext{\vspace{-10mm}}
\begin{document}

\title{TALU: A Reconfigurable Tiled-ALU Grid for \\RISC-V}

\author{
\IEEEauthorblockN{
Ali Emre Oztas\IEEEauthorrefmark{1}*,
Can Erim Karsal\IEEEauthorrefmark{2}*,
Mahir Demir\IEEEauthorrefmark{1}*,
Mikel Luj{\'a}n\IEEEauthorrefmark{1}
}
\IEEEauthorblockA{
\IEEEauthorrefmark{1}University of Manchester, UK\\
\IEEEauthorrefmark{2}Technical University of Munich, Germany\\
\{aliemre.oztas, mahir.demir\}@manchester.ac.uk,
can.karsal@tum.de
}
\thanks{* These authors contributed equally to this work.}
}

\maketitle

\thispagestyle{plain}
\footnotetext{This work has been submitted to the IEEE for possible publication. Copyright may be transferred without notice, after which this version may no longer be accessible.}

\begin{abstract}
Lane-based RISC-V vector processors exploit data parallelism well but pay a
recurring cost for two common patterns: reductions, which require a separate
tree pass, and deep dependent expression chains, which cannot be fused and are
issued as serial vector instructions. We present \textbf{TALU}, a 5-stage
RISC-V core whose execute stage is an $N{\times}N$ grid of registered ALU tiles
instead of a single ALU. The grid is a reconfigurable, feed-forward spatial
fabric: tiles route operands only from non-backward neighbours, and added
diagonal forwarding lets a reduction collapse spatially. A minimal custom extension
configures the per-tile operation, routing, and per-tile constants, and streams
operands through the array. On a single fabric TALU runs ordinary scalar code,
single deep-path expressions, and multi-lane vector kernels. On the benchmark
suite of the RISC-V$^2$ vector processor, TALU is $1.8$--$5.5\times$ faster
than RISC-V$^2$ and $4.5$--$15.5\times$ faster than
Vicuna at equal lane count, with a reduction cost
(dot$\div$add) of $1.05\times$ versus $1.96\times$ and $1.26\times$ for the
vector cores. A retimed implementation places and routes at \textbf{1.03\,GHz}
in 45\,nm in 0.30\,mm$^2$, half the area of RISC-V$^2$.
\end{abstract}

\begin{IEEEkeywords}
RISC-V, spatial datapath, CGRA, vector reduction, operation chaining, ASIC.
\end{IEEEkeywords}

\section{Introduction}
Vector architectures combine high programmability with high throughput and are the workhorse for data-parallel workloads; examples on RISC-V~\cite{rvvspec,riscv2,vicuna,ara,spatz}. Despite these achievements, the lane-based, time-multiplexed organisation is inefficient for two common computational patterns part of DSP and Machine Learning kernels. The first one is \textit{fused reductions} (e.g., dot product, $L_2$
norm, sum), which are not calculated by the vector lanes themselves. These operations require a dedicated
hardware reduction tree, whose latency grows logarithmically and a separate pass over the data~\cite{riscv2}. The second is deep dependent expression chains such as $(a{+}b)^n$, such as the form of the polynomial kernel $(\mathbf{u}\!\cdot\!\mathbf{v}+1)^{d}$ of support vector
machines~\cite{cortes1995}. These chains cannot be fused with existing vector instructions, and thus, a RISC-V vector processor has to
issue each dependent operation as its own vector instruction.

Spatial datapaths address both. Arranging operators in space enables a fused reduction to be
an operate and forwarding combination over a grid and helps a dependent chain to become a pipeline. Coarse-grained reconfigurable arrays (CGRAs) and in-pipeline reconfigurable units have exploited this previously~\cite{dyser,cgra_survey}. Our contribution is therefore neither spatial
computing, nor in-pipeline integration, but a specific co-design and its
measured consequences in a RISC-V core. Three choices define it: the column index orders the
dataflow (data only moves to higher columns), the diagonals are the only block-aligned
cross-row path in a streaming pass, and one combined scalar/vector register
path feeds a fabric that also runs ordinary RV32I + MUL. We
propose \textbf{TALU} (Tiled ALU), whose execute stage is an $N{\times}N$
grid of registered ALU tiles. For ordinary scalar code, tile $(0,0)$
also acts as the scalar ALU. Thus, the entire system
generates one fabric that serves three execution modes: scalar, single-deep-path, and
multi-path. Our contributions can be summarized as follows:

\begin{figure}[t]
\centering
\includegraphics[width=\columnwidth]{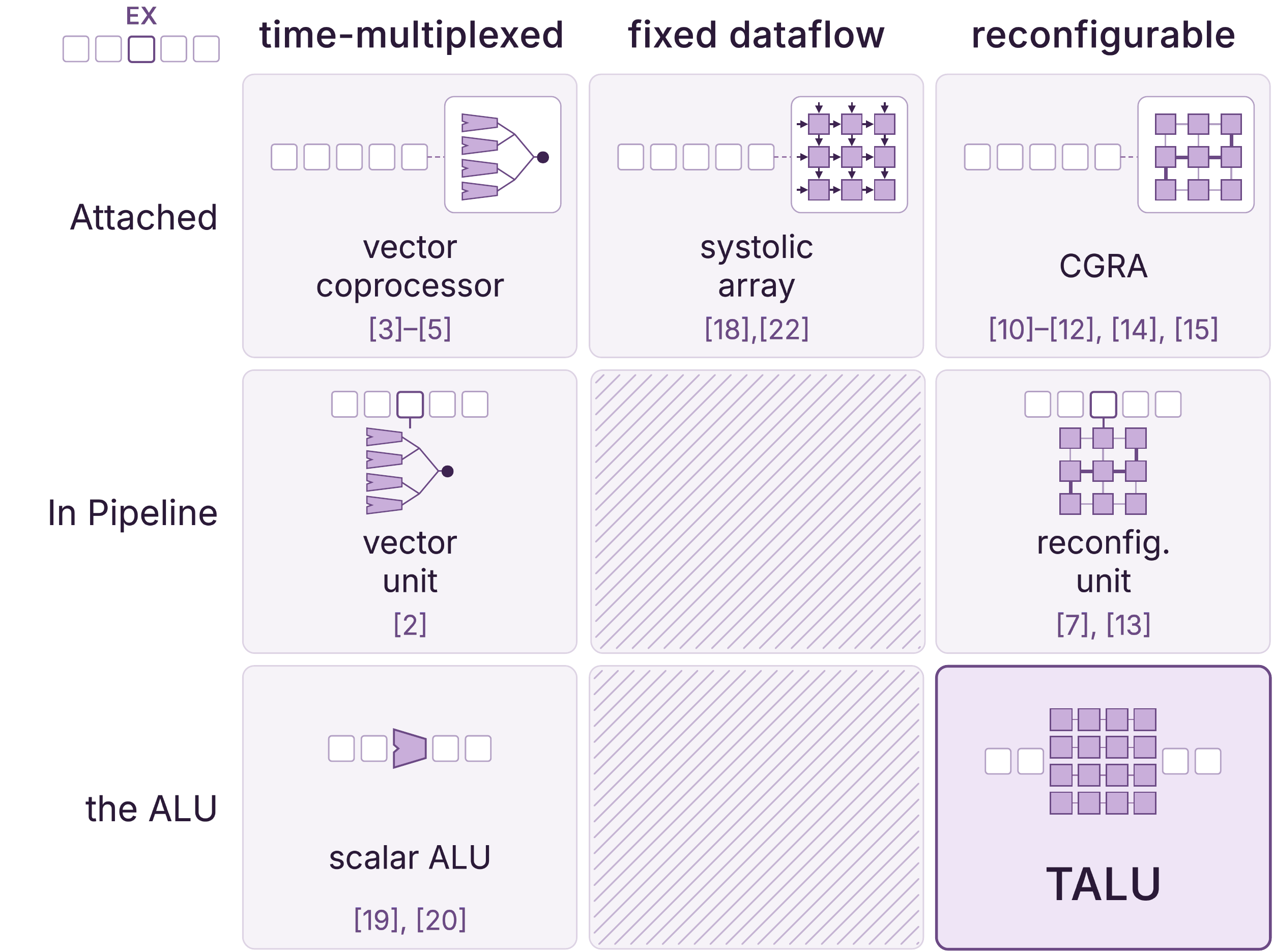}
\caption{Spatial datapaths by how the array computes (columns) and where it
sits (rows); only TALU's fabric \emph{is} the execute ALU.}
\label{fig:pos}
\end{figure}

\begin{itemize}

\item The reconfigurable grid replaces the
single execute ALU rather than sitting beside it as a coprocessor, sharing the
scalar forwarding network and the (scalar + vector) register files.
\item A column-ordered dataflow whose \emph{diagonal} (NW/SW) links are
the only block-aligned cross-row path in a streaming pass, so a reduction
collapses inside the fabric rather than in a separate pass over the data;
\emph{segmented} (per-block) reductions, which a conventional adder tree
cannot express, run $14.4\times$ faster than a scalar loop (Sec.~\ref{sec:eval}).
\item A minimal extension exposing the
three modes, with per-tile \emph{constants} and a stream tap that fuse a
multiply-accumulate into one pass (saxpy in one pass, a $K$-tap FIR in $K$);
an evaluation on the RISC-V$^2$
paper's own benchmark suite against \emph{two} independent vector cores; and
a 45\,nm implementation that closes
place-and-route at \textbf{1.03\,GHz in 0.30\,mm$^2$} of die, half the area of
the vector baseline.
\end{itemize}

\begin{figure}[t]
    \centering
    \includegraphics[width=\columnwidth]{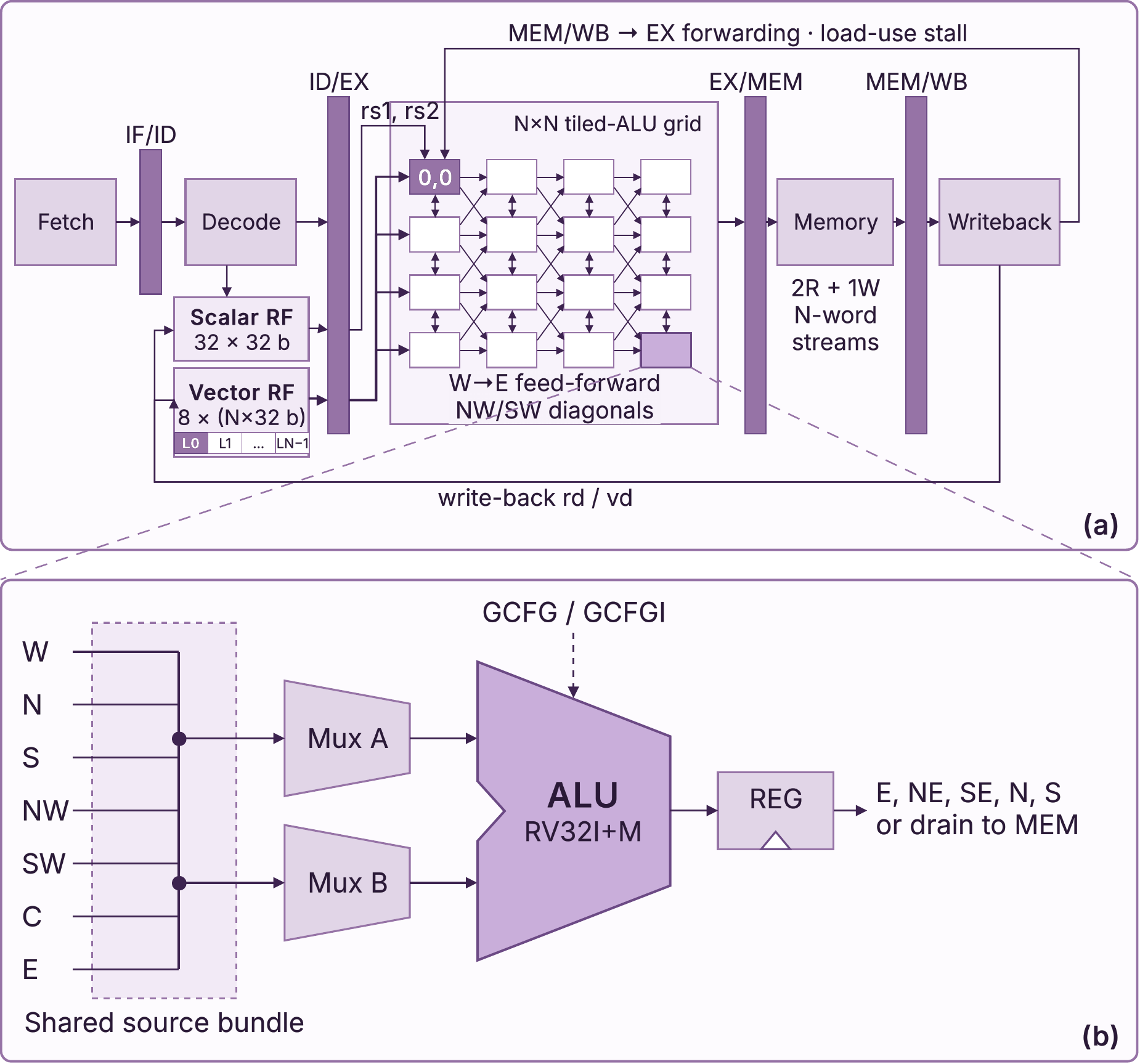}    \caption{TALU architecture. (a)~$N{\times}N$ grid in the EX stage, fed by the
    scalar and vector register files; tile $(0,0)$ is the scalar ALU. (b)~One tile.}
    \label{fig:arch}
\end{figure}
\section{Related Work}
\textbf{CGRAs and in-pipeline reconfigurable datapaths.} Standalone
coarse-grained reconfigurable arrays span ADRES~\cite{adres},
HyCUBE~\cite{hycube} and Plasticine's parallel-pattern fabric~\cite{plasticine},
and have reached commercial scale in the AMD/Xilinx Versal AI Engine tile
array~\cite{versal}; all are accelerators
\emph{beside} a host core. Closer to us, DySER places a circuit-switched array
of heterogeneous functional units in the execute stage of a
pipeline~\cite{dyser}, and Chimaera~\cite{chimaera},
PipeRench~\cite{piperench} and Garp~\cite{garp} couple reconfigurable units
tightly to a CPU. None makes the array \emph{the} ALU
(Fig.~\ref{fig:pos}).

\textbf{RISC-V vector processors.} RISC-V$^2$~\cite{riscv2} is a throughput-optimized
vector core with dynamic register grouping and a hardware reduction
tree; Vicuna~\cite{vicuna} is a
timing-predictable embedded \texttt{Zve32x} coprocessor. We compare against
both. Ara~\cite{ara} and Spatz~\cite{spatz} scale the same lane
organization~\cite{rvvspec} up and down, and the
programmability-vs.-efficiency structure of such data-parallel machines is
well studied~\cite{lee2011,kozyrakis2003}.

\textbf{Systolic arrays and streaming extensions.} Fixed-dataflow MAC grids
such as the TPU~\cite{tpu} also pipeline data spatially, but hardwire the
operation and the reduction direction for GEMM. Stream semantic registers~\cite{ssr} hide load/store and loop
overhead to keep a scalar FPU fully utilized; \texttt{GFLOW} shares that goal
but streams through a spatial array. A recent scalar-chaining ISA
extension~\cite{chaining} fuses dependent operation chains in the issue stage,
whereas TALU fuses them in \emph{space}. Vector-permutation
support~\cite{perm} likewise targets patterns that plain lanes express poorly.
Closest among \emph{reuse}-oriented designs is VSA~\cite{vsa}, which remaps the
lanes of an existing RVV unit into a GEMM systolic array for under $0.1\%$
extra area. TALU instead presumes no vector unit and targets reductions and
dependent chains across a general kernel set; VSA is the better answer when a
VPU is already present and the workload is GEMM.

\begin{figure}[t]
\centering
\includegraphics[width=0.98\columnwidth]{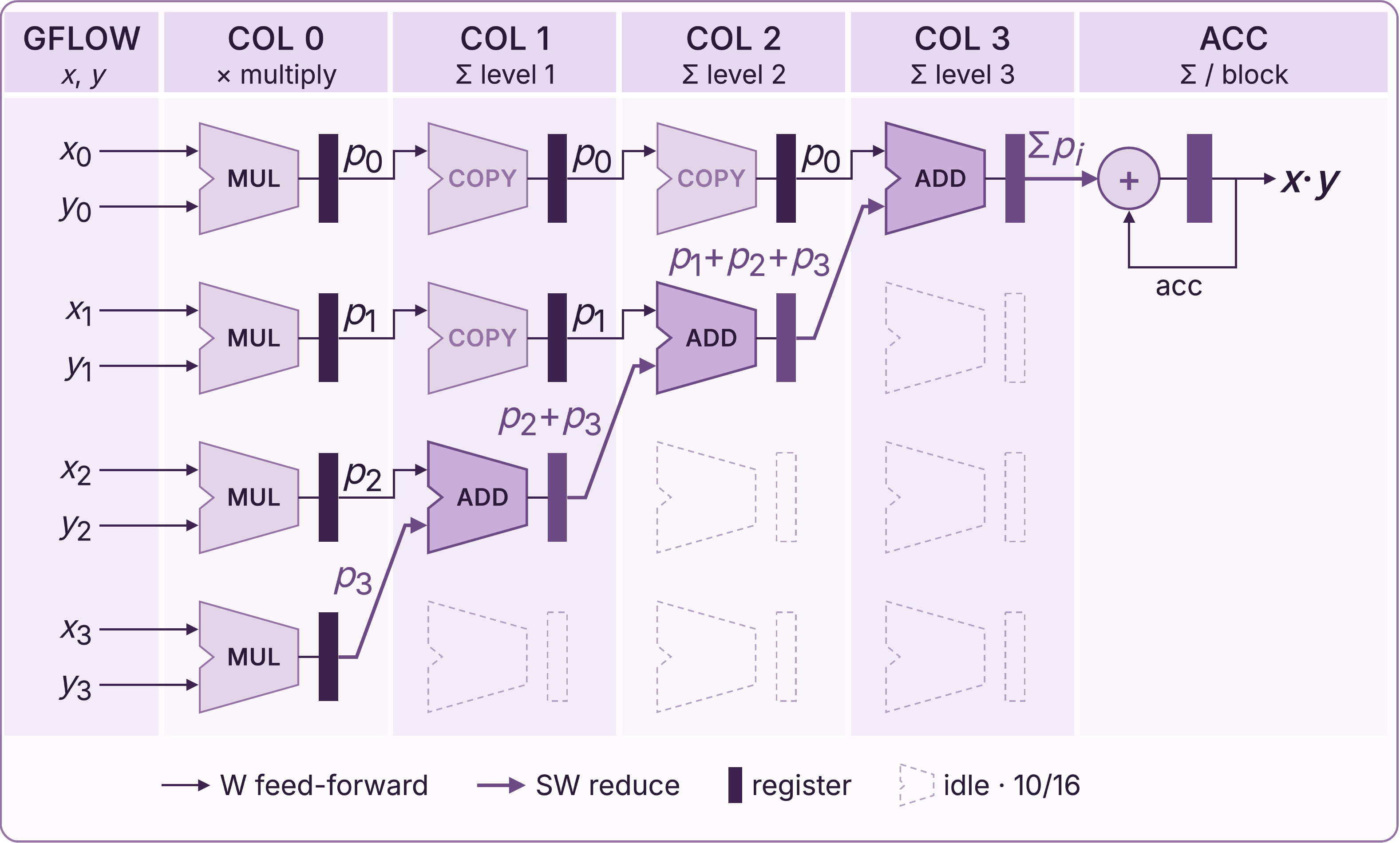}
\caption{Dot product on the grid ($N{=}4$): column~0 multiplies the two
streams; an \texttt{ADD(W,SW)} chain climbs one row per column to the corner.}
\label{fig:map}
\end{figure}

\section{TALU Architecture}\label{sec:arch}
TALU is a classic 5-stage pipeline; the EX stage holds the
$N{\times}N$ grid (Fig.~\ref{fig:arch}(a)). Each \emph{tile} $(r,c)$ contains a
combinational RV32I + MUL ALU and one output register
(Fig.~\ref{fig:arch}(b)). Operands are selected, independently per input,
from non-backward neighbours only. No tile reads a higher column, so the
dataflow is acyclic across columns; the same-column N/S links read a
neighbour's registered output, so they never form a combinational loop (no
kernel here uses them). Column~0 is fed by injected operands and any
perimeter (and input-column) tile may drain forward to MEM.

\textbf{Latency vs.\ throughput.} Because every tile is registered, a path of $k$
tiles has \emph{latency} $k$ cycles but does not lower the clock period---each
tile is its own pipeline stage. Equal-length paths therefore drain together.

\textbf{Diagonal forwarding for in-pass reduction.} In a streaming pass each
column holds one block, so a tile's W, NW and SW inputs all see the same block.
The NW/SW diagonals therefore let a tile add two adjacent lanes of that block,
and an $N$-element reduction becomes a diagonal chain that collapses to the
top-right corner while the next blocks stream in, without stalls. For the
dot product of Fig.~\ref{fig:map}, ten configuration instructions set up the
grid once and a single streaming instruction then processes both vectors.
Because the corner emits one sum per block, the same
chain also gives \emph{segmented} (per-block) reductions, which an accumulator
over the drained lanes cannot produce (Sec.~\ref{sec:eval}).

\textbf{Register files.} A scalar 32$\times$32 file feeds tile $(0,0)$; an
$N$-lane vector file (each register $N{\times}32$\,b) feeds the grid. The same
vector registers serve scalar use (lane-0 slice) and multi-path use (all lanes).
The data memory exposes a scalar word/half/byte port and wide $N$-word ports:
two read streams plus one write stream, a peak of $3N$ words/cycle in
streaming mode (measured in Sec.~\ref{sec:eval}).
\subsection{Custom ISA and Execution Modes}
A small custom extension drives the fabric:
\texttt{GCFG} configures one tile (operation---the RV32I + MUL set plus
\texttt{MIN}/\texttt{MAX} and copies---two operand sources, and
drain); \texttt{GCFGI} loads a scalar register into a per-tile \emph{constant}
register (source \texttt{C}: coefficients, taps);
\texttt{GSTART} runs the configured grid on register-resident operands; and
\texttt{GFLOW} \emph{streams} a memory region through the grid, injecting one
$N$-wide block and draining one per cycle (path latency derived in hardware
from the drain map). A tile may also tap the second stream at \emph{any}
column (source \texttt{E}): hardware derives the tap column and delays that
stream's fetch to match, so a multiply-accumulate fuses into a single pass
(col\,0 \texttt{MUL(W,C)} $\to$ col\,1 \texttt{ADD(W,E)}). One fabric
supports three modes:
(i)~\emph{scalar} RV32I + MUL  via tile $(0,0)$'s combinational tap (with forwarding
and load-use stalls on that path only);
(ii)~a \emph{single deep path} that fuses a dependent expression across columns;
(iii)~\emph{multi-path}, $N$ equal-length lanes draining together. Unlike the scalar mode, modes (ii) and (iii) do not check for hazards between instructions, since the streaming model guarantees that none occur.

\subsection{Compiling Kernels onto the Fabric}\label{sec:compile}
A prototype LLVM pass maps annotated single-pass C loops. It extracts the
loop-body expression DAG, classifies it as a map or a reduction, and places it
mechanically: since every tile is registered, a node at DAG depth $d$ goes to
column $d{-}1$, so a dependent chain becomes a column pipeline; constants use
the tile's constant register, and a reduction appends the diagonal collapse of
Fig.~\ref{fig:map}. Of twelve annotated loops, the eight maps (vvadd, vmul,
vsub, vmin, vmax, scale, saxpy, $(a{+}b)^8$) and three reductions (dot, sum,
sum of squares) compile and match the same C run natively; the seven with a
hand-written counterpart are \emph{tile-for-tile identical} to it, with the
same cycle count. The twelfth, a three-input median, is rejected with a
diagnostic, as \texttt{GFLOW} supplies two streams. Multi-pass kernels (FIR,
the 4-pass median, SAD) are still mapped by hand.

\section{Evaluation}\label{sec:eval}
TALU RTL is simulated cycle-accurately (Icarus Verilog) and a custom assembler assembles kernels. Every module has a self-checking testbench,
every kernel is checked against hand-computed results on an 8-16 element
input, and the compiled kernels are checked word for
word at an application vector length (AVL) of 64 against the same C source run natively. Baselines run in their own
testbenches: RISC-V$^2$~\cite{riscv2} via its Verilator simulator (4 lanes, ideal
memory), and Vicuna~\cite{vicuna} via Verilator (\texttt{Zve32x} on an Ibex host,
$\mathrm{VREG\_W}{=}128$, $\mathrm{VMEM\_W}{=}64$, LMUL\,$m8$). All process the same element count with four lanes/rows
of 32-bit elements. TALU uses $N{=}4$ with an ideal wide scratchpad; we report
the configuration-independent \emph{ratio} metrics as the robust results.
Unless stated otherwise, every TALU number is the \emph{retimed 1\,GHz} RTL of Sec.~\ref{sec:phys}; the
baseline (one combinational tile stage) is carried alongside in
Table~\ref{tab:cyc} so the cost of retiming is visible.

\textbf{Benchmarks.} We adopt the kernel suite of the RISC-V$^2$
paper~\cite{riscv2}: FIR, dot product, median, multiply, vvadd, saxpy, motion
estimation, perceptron (we add an array-mean ``average'' kernel and the
deep-chain $(a{+}b)^8$). The RISC-V$^2$ \emph{fir} and \emph{saxpy} kernels are
that repository's own examples, unmodified; multiply/average/perceptron are
direct transliterations into its ISA, which lacks
\texttt{vmin}/\texttt{vmax}, so median and motion estimation are not
expressible on it (Vicuna's \texttt{Zve32x} covers both). The suite's
CNN/DCNN entries are whole-network MNIST benchmarks that are not reproducible
on the public platforms. On TALU, single-op, reduce and saxpy kernels are single
\texttt{GFLOW} passes (Fig.~\ref{fig:map}); 5-tap FIR runs one MAC pass per
tap (accumulator in place); median is a 4-pass min/max network; motion
estimation (SAD) is two reduce passes.

\textbf{Results.} Table~\ref{tab:cyc} reports cycles for 1024 elements; TALU
is fastest on every kernel against both vector cores. Three observations.
\emph{(i) Map and reduce run at the same throughput}: vvadd, dot product,
perceptron and $(a{+}b)^8$ land within $8\%$ of one another, because reduction
and chaining add only a fixed pipeline fill, whereas both vector cores pay
substantially more for either (the $\div$add rows). \emph{(ii) Spatial fusion
makes saxpy cost the same as vvadd}, where the vector cores pay
$1.4$--$2.2\times$ their vvadd for the extra dependent instruction pass.
\emph{(iii) Multi-pass kernels win least}: FIR, median and SAD pay one full
streaming pass per MAC or min/max stage, bounded by the
two-operation-per-pass ceiling of the current extension. Retiming for 1\,GHz
costs only $1.8\%$ of cycles over the baseline RTL, because the deeper fill is
paid once per \texttt{GFLOW} pass rather than per element ($+14\%$ at
$\mathrm{AVL}{=}64$, $+0.5\%$ at 4096).

\textbf{From cycles to time.} A cycle comparison settles execution time only
at equal clocks. TALU closes place-and-route at 1.03\,GHz and RISC-V$^2$
reports 1\,GHz, so the ratios of Table~\ref{tab:cyc} carry over to time:
$3.2\times$ geomean in cycles, $3.3\times$ in time, although the two clocks
are not corner-matched (Sec.~\ref{sec:phys}). Without retiming, the
baseline tile closes place-and-route at only 674\,MHz, which would cut this
advantage to $2.2\times$.

\textbf{Cycle model and vector length.} TALU's streaming cycle count is well
approximated by $C \approx S + \mathrm{AVL}/N + L$, where $S$ is a one-time
setup (${\sim}20$ cycles including configuration) and $L$ the drained path
depth, which is why reduction and chain kernels exceed element-wise ones only
by their fill. The fill amortizes with vector length: sweeping
$\mathrm{AVL}=256/1024/4096$, TALU's dot$\div$add is $1.17/1.05/1.01$
(${\to}1$) while RISC-V$^2$'s is $1.85/1.96/1.99$ (${\to}2$), because its
reduction-tree pass scales \emph{with} the data; TALU's dot-product speedup
therefore grows with size ($3.5\times$, $4.5\times$, $4.9\times$).

\begin{table}[t]
\caption{Cycles for 1024 elements, 4 lanes/rows, RISC-V$^2$
suite~\cite{riscv2}. \emph{base}: one tile stage; \emph{1\,GHz}: retimed RTL,
used throughout.}
\label{tab:cyc}
\centering
\scriptsize
\setlength{\tabcolsep}{0pt}
\begin{tabular*}{\columnwidth}{@{\extracolsep{\fill}}l rr rr rr@{}}
\toprule
 & \multicolumn{2}{c}{TALU} & & & \multicolumn{2}{c}{speedup vs.} \\
\cmidrule(lr){2-3}\cmidrule(l){6-7}
kernel & base & \textbf{1\,GHz} & RISC-V$^2$ & Vicuna & RV$^2$ & Vic. \\
\midrule
vvadd, multiply   & 270  & \textbf{273}  & 661  & 2037 & $2.4\times$ & $7.5\times$ \\
saxpy             & 290  & \textbf{296}  & 1422 & 2879 & $4.8\times$ & $9.7\times$ \\
dot product       & 280  & \textbf{287}  & 1297 & 2560 & $4.5\times$ & $8.9\times$ \\
FIR (5-tap)       & 1363 & \textbf{1389} & 2825 & 6613 & $2.0\times$ & $4.8\times$ \\
median (3-wide)   & 1071 & \textbf{1080} & n/a  & 4863 & ---         & $4.5\times$ \\
motion est.\ (SAD)& 554  & \textbf{567}  & n/a  & 3592 & ---         & $6.3\times$ \\
perceptron        & 286  & \textbf{293}  & 1613 & 2692 & $5.5\times$ & $9.2\times$ \\
$(a{+}b)^8$       & 285  & \textbf{291}  & 1353 & 4523 & $4.6\times$ & $15.5\times$ \\
average           & 285  & \textbf{292}  & 526  & 1633 & $1.8\times$ & $5.6\times$ \\
\midrule
total / geomean   & 4954 & \textbf{5041} &      &      & $\mathbf{3.2\times}$ & $\mathbf{7.4\times}$ \\
dot $\div$ add    & 1.04 & \textbf{1.05} & 1.96 & 1.26 & & \\
$(a{+}b)^8\div$ add & 1.06 & \textbf{1.07} & 2.05 & 2.22 & & \\
\bottomrule
\end{tabular*}
\end{table}

\textbf{Deep chaining vs.\ scalar.} As a sanity control, $(a{+}b)^8$ on
TALU's own fully-forwarded scalar path is $77\times$ slower than on the grid
(partly because the retimed scalar \texttt{MUL} takes 3 cycles), so the win is the single-issue \emph{issue bound}, not missing forwarding.
Ablating grid width attributes it: at $N{=}1$, one ALU but still streaming,
vvadd already runs $12.8\times$ faster than the scalar loop, and widening to
$N{=}4$ adds only $3.8\times$. \emph{The dominant effect is removing the
per-element issue bottleneck}; the array then multiplies it.

\begin{figure}[t]
\centering
\includegraphics[width=\columnwidth]{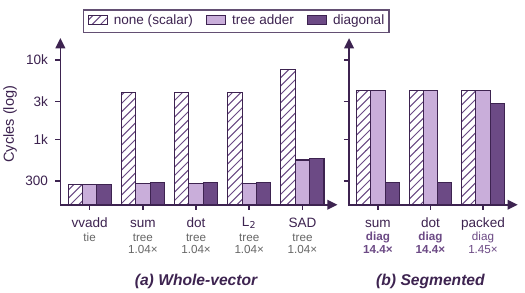}
\caption{Reduction cycles ($\mathrm{AVL}{=}1024$) with no reduction hardware,
the tree adder, or the diagonal; best mechanism below each kernel.}
\label{fig:red}
\end{figure}

\textbf{What the diagonal actually buys.} \texttt{GFLOW} also has a
conventional $N$-input adder tree that sums the drained lanes into one
accumulator. Fig.~\ref{fig:red} compares it with the diagonal and with no
reduction hardware, where a scalar loop sums the stored lanes afterwards, an
order of magnitude slower. For reductions to a single value, the adder is
slightly faster ($1.04\times$), as the diagonal pays 12 extra fill cycles per pass.
For \emph{segmented} reductions (one result per block, e.g., batched inner
products), only the diagonal works: the adder yields a single total, so the
scalar loop returns and the diagonal wins by $14.4\times$ ($1.45\times$ if
its strided output must be packed). The diagonals cost only $3.5\%$ of cell
area.

\textbf{Memory model.} Streaming $N$ elements/cycle needs $3N$ words/cycle
($2N$ read, $N$ write; 12 at $N{=}4$), which a conventional banked TCDM
(tightly coupled data memory) provides without multi-ported SRAM: a
contemporary RISC-V cluster sustains 32 words/cycle across 32
banks~\cite{zerostall}. Replaying the vector-port access traces through a
16-bank TCDM model brings every kernel to within 3 cycles of its ideal count,
provided concurrently read streams are offset by a multiple of $N$ words;
with power-of-two-aligned arrays they collide on the same banks (up to
$1.95\times$). Below peak bandwidth, every advantage over the best baseline
holds down to $1.5N$ words/cycle.

\section{Physical Implementation}\label{sec:phys}
We implemented the retimed TALU in 45\,nm (FreePDK45~\cite{freepdk45},
NanGate cells) with Synopsys DC and Cadence Innovus through place-and-route,
excluding memories and signing off setup at the typical corner
(Table~\ref{tab:hw}). At $N{=}4$ it closes the 1.0\,ns target with a DRC- and
connectivity-clean layout. Retiming is what makes this possible: the baseline
tile, with a flat 32-bit multiplier, closes place-and-route at only
674\,MHz (1.5\,ns target, same corner), while pipelining the multiplier inside
its carry-save tree reaches 1032\,MHz ($1.53\times$) for $54\%$ more flops and
the $1.8\%$ extra cycles of Table~\ref{tab:cyc}.

\textbf{Scaling with $N$.} Table~\ref{tab:hw} sweeps the grid from
$N{=}1$ to~8. The clock stays flat, so grid-control fan-out does not limit
scaling, and element-wise throughput grows almost linearly with~$N$.
Reductions keep pace: their only overhead is a fill of $N$ columns, paid once
per pass. Area, however, grows with the $N^2$ tiles, so throughput per area is
best at small $N$; the extra columns of a larger grid buy path depth for fused
chains rather than width. With the multiplier pipelined,
removing multipliers no longer raises the clock: a column-0-only fabric, which
still runs nine of the ten kernels (all but $(a{+}b)^8$), simply saves a third
of the cell area.

Against RISC-V$^2$, TALU occupies half the area. The clock and power figures
are not like-for-like: RISC-V$^2$ reports 1\,GHz from synthesis at the
worst-case corner and a lower supply, whereas ours are post-route at the
typical corner with a 0.1-activity power estimate. TALU's power is higher, and
energy per element is the main open question.

\begin{table}[h]
\caption{Hardware implementation comparison at 45\,nm.}
\label{tab:hw}
\centering
\scriptsize
\setlength{\tabcolsep}{0pt}
\begin{tabular*}{\columnwidth}{@{\extracolsep{\fill}}l rr r r rr@{}}
\toprule
 & \multicolumn{2}{c}{area (mm$^2$)} & $f_{\max}$ & power
 & \multicolumn{2}{c}{$\mathrm{AVL}{=}1024$} \\
\cmidrule(lr){2-3}\cmidrule(l){6-7}
design & cell & die & (MHz) & (mW) & add el/cyc & dot$\div$add \\
\midrule
TALU, $N{=}1$          & 0.033 & 0.071 & 1062 & 36  & 0.98 & 1.00 \\
TALU, $N{=}2$          & 0.061 & 0.119 & 1044 & 66  & 1.94 & 1.01 \\
\textbf{TALU, $\boldsymbol{N{=}4}$} & \textbf{0.166} & \textbf{0.300} & \textbf{1032} & \textbf{178} & \textbf{3.75} & \textbf{1.05} \\
TALU, $N{=}8$          & 0.605 & 1.221 & 1010 & 652 & 6.87 & 1.30 \\
\midrule
$N{=}4$, col-0 MUL only & 0.110 & 0.211 & 1059 & 109 & 3.75 & 1.05 \\
$N{=}4$, no MUL         & 0.088 & 0.180 & 1041 & 81  & 3.75 & -- \\
\midrule
RISC-V$^2$, 4 lanes     & \multicolumn{2}{c}{0.61} & 1000 & 43.9 & -- & -- \\
\bottomrule
\end{tabular*}
\end{table}

\section{Conclusion}
TALU turns the RISC-V execute stage into a reconfigurable, feed-forward ALU
grid whose diagonal links fold reductions into the streaming pass and whose
columns fuse dependent chains. On the RISC-V$^2$ benchmark suite it is faster
on every kernel than two vector cores with low-cost reductions, and a retimed
implementation closes 1.03\,GHz in 45\,nm in half the area of RISC-V$^2$. Most
of the gain comes from removing the per-element issue bound, which the array
then multiplies; the diagonals add segmented reductions an accumulator cannot
express. The RTL, kernels, compiler and evaluation scripts are available at
\url{https://github.com/circuitmaster/TALU-a-Tiled-ALU-Grid-in-the-RISC-V-Execute-Stage}.

\bibliographystyle{IEEEtran}
\bibliography{bib}

\end{document}